\documentclass[journal,onecolumn]{IEEEtran}
\usepackage{cite}
\usepackage{graphicx}
\usepackage[cmex10]{amsmath}
\usepackage{amssymb}
\usepackage{subfigure}
\usepackage{subfloat} 
\usepackage{graphics}
\usepackage{bigints}
\usepackage{amsthm}
\usepackage{xcolor}
\usepackage{epsfig,psfrag}
\usepackage{epstopdf}
\usepackage{setspace}
\usepackage{multirow}
\usepackage{amssymb,dsfont,graphicx}
\usepackage{cite}
\usepackage{tablefootnote}

\newcommand{\tr}{\text{\,tr\,}}
\newcommand{\diag}{\text{\,diag\,}}

\newcommand{\MG}[6]{
G^{3,1}_{2,3} \left(
\begin{array} {c}
#1,\, #2 \\ #3,\, #4,\, #5 \end{array}\Bigg|\, #6 \right)}

\newcommand{\MGG}[8]{
G^{3,2}_{3,4} \left(
\begin{array} {c}
#1,\, #2,\, #3 \\ #4,\, #5,\, #6,\, #7 \end{array}\Bigg|\, #8 \right)}

\begin{document}

\title{On the Exact and Approximate Eigenvalue Distribution for Sum of Wishart Matrices}

\author{Santosh Kumar, Gabriel Fernando Pivaro, Gustavo Fraidenraich,  and Claudio Ferreira Dias 
\thanks{Santosh Kumar is with 
the Department of Physics, Shiv Nadar University, India. G. F. Pivaro, G. Fraidenraich, and C. F. Dias are with the Department of Communications, State University of Campinas (Unicamp), Brazil.}

}
\date{\vspace{-5ex}}

\maketitle

\doublespacing

\begin{abstract}

The sum of Wishart matrices has an important role in multiuser communication employing multiantenna elements, such as multiple-input multiple-output (MIMO) multiple access channel (MAC), MIMO Relay channel, and other multiuser channels where the mathematical model is best described using random matrices.

In this paper, the distribution of linear combination of complex Wishart distributed matrices has been studied. We present a new closed form expression for the marginal distribution of the eigenvalues of a weighted sum of $K$ complex central Wishart matrices having covariance matrices proportional to the identity matrix. The expression is general and allows for any set of linear coefficients.

As an application example, we have used the marginal distribution expression to obtain the ergodic sum-rate capacity for the MIMO-MAC network, and the cut-set upper bound for the MIMO-Relay case, both as closed form expressions. 
	
We also present a very simple expression to approximate the sum of Wishart matrices by one equivalent Wishart matrix. All of our results are validated by means of Monte Carlo simulations. As expected, the agreement between the exact eigenvalue distribution and simulations is perfect, whereas for the approximate solution the difference is indistinguishable. 

\end{abstract}

\begin {IEEEkeywords}
Sum of Wishart matrices, eigenvalue distribution, multiple-input multiple-output, ergodic sum capacity,  Meijer-G function.
\end{IEEEkeywords}

\section{Introduction}

\subsection{Random matrices and MIMO single-user relation}

\IEEEPARstart{R}{}andom matrix theory has evolved into a truly multidisciplinary subject with its applications in fields as varied as communication theory, quantum transport, quantum chromodynamics, quantum information theory, string theory, econophysics, number theory, etc. \cite{Akemann2011}. It is possible to represent the \emph{operators} relevant to study physical systems in matrix form and use its properties to tackle difficult problems. Communication theory is one of the prominent areas on which random matrix theory has had huge impact. Random matrices gained special attention in wireless communication field after the works of Winters\cite{winters1987capacity}, Foschini\cite{foschini1996layered}, and Telatar\cite{telatar1999capacity}. They have shown that the use of multiple antennas could enhance capacity in systems with limited bandwidth. In all cases, mathematical tools for matrices were employed for the analysis.

In Telatar's paper \cite{telatar1999capacity}, the multiple-input multiple-output (MIMO) point-to-point ergodic channel capacity has been found. He has shown that instead of dealing with the joint probability density function of a Wishart distributed matrix \cite{WISHART1928}, which is not easy to handle even for small dimensions, it suffices to use the joint eigenvalue probability density function given by James \cite{James1960}. Such a simplification is possible in view of the unitarily-invariant nature of the channel capacity and other metrics which are usually employed to characterize the MIMO systems. Telatar's pioneering work was the first one to establish a connection between MIMO communication and random matrix theory. Since then, there has been a great deal of interest in exploring and comprehending the properties of Wishart matrices. 

As a special case of Hermitian matrices, Wishart matrices arise in scenarios where MIMO systems are subject to Rician or Rayleigh fading. As indicated above, the performance of MIMO systems can be statistically predicted with the aid of eigenvalues distribution of Wishart matrices \cite{telatar1999capacity}, \cite{Zanella2009}. For example, the channel matrix in a MIMO system relates to a Wishart matrix, whose eigenvalue statistics then leads to the knowledge of the ergodic capacity of the MIMO channel \cite{telatar1999capacity}. On the other hand, the distribution of the largest and smallest eigenvalue can be used to analyze the performance of MIMO maximal ratio combining systems and MIMO antenna selection techniques, respectively \cite{Zanella2009}. In \cite{McKay2008a}, the authors have shown that the symbol error rate (SER) performance of MIMO systems
employing multichannel beamforming in arbitrary-rank Ricean channels is dominated by the subchannel SER corresponding to the minimum channel singular value. Their results are based on marginal ordered eigenvalue distributions of complex noncentral Wishart matrices.  Since in the slow fading scenario it is not feasible to determine the ergodic capacity, a metric denominated outage probability is required to evaluate the system performance \cite{tse2005fundamentals}. The outage probability is related to the cumulative eigenvalue distribution function of a Wishart matrix \cite{telatar1999capacity}, \cite{McKay2008a}, \cite{Wang2004}. Due to physical nature of wireless channel and all possible arrangements for antennas arrays, different types of Wishart matrices have been studied, such as central and noncentral, associated with Rayleigh and Rician fading, respectively; uncorrelated, semi-correlated, and double-correlated, associated with the antenna correlation at the transmitter side and at the receiver side \cite{Ordoez2009}, \cite{Smith2003}.

\subsection{Extension for MIMO multiuser case}

All the works previously mentioned are concerned with the MIMO single-user channel where the majority of the problems are already solved or at least well understood \cite{Goldsmith2003}. However, for the MIMO multiuser scenario there exists many open problems, such as the general capacity for the MIMO Relay channel.

In a wireless multiuser channel, we are generally more concerned in the overall information rate (capacity) of the system than the individual user rates as in the single user channel \cite[Ch. 15]{cover2006elements}. In this way, we could define metrics associated with the joint users performance. We have, for example, symmetric capacity and sum capacity. The former is the maximum common rate at which both users can simultaneously reliably communicate; the latter is the maximum total throughput that can be achieved \cite[pg. 230]{tse2005fundamentals} and can be seen as a constraint that limits the individual rates of each user. Since sum capacity reflects an overall system performance, this metric is of great interest from analytical and practical points of view.

As can be inferred from the sum capacity name, to evaluate this metric we have to add up the rates of each user. This operation leads to a summation of Wishart matrices associated with each one of the MIMO channels involved in the multiuser system. For example, this situation occurs in two well-known multiuser channels: (i) MIMO multiple access channel (MAC), where $K$ users with multiple transmit antennas communicate with one destination also with multiple receiving antennas \cite{Hochwald02}; and (ii) MIMO Relay channel, where a MIMO transmitter communicates with a MIMO receiver with the help of a MIMO relay \cite{Host-Madsen2005}. For MIMO MAC the sum capacity is a desired metric on performance \cite{Goldsmith2003}. For MIMO Relay, the sum capacity is used to determine the cut-set upper bound on the channel capacity \cite{Host-Madsen2005}.

\subsection{On the Paper Contribution} 

Based on our discussion in the preceding section about the sum capacity, hereinafter, we will analyze this metric under fast-fading Rayleigh distribution. Therefore, our aim is to determine the \emph{ergodic sum capacity} for MIMO multiuser scenario. Our idea is to use the framework identical to that of a single-user case. It means that we wish to obtain the ergodic sum capacity using the marginal eigenvalue distribution of the sum of Wishart matrices. 

For a single-user case, the probability density function of the eigenvalues of a Wishart matrix was given in \cite{James1960}, and since then many advances have been achieved for the most variate cases of Wishart distributions. More recently, results on the product of rectangular random matrices have appeared in \cite{Akemann2013} and \cite{Akemann20132} and the authors have investigated ergodic mutual information in MIMO communication channel with multifold scattering. By contrast, the progress for the eigenvalues distribution of the sum of Wishart matrices has not been going at the same pace.

Although the most well-known result for the sum of Wishart matrices dates back to 1960's, it is valid only for the specific case where all matrices have the same covariance matrix \cite{anderson1958}; not much is known for the general case of arbitrary covariance matrices. In \cite{Tan1983}, the authors have considered linear combination of central Wishart matrices with positive coefficients. They have proposed approximating the distributions of the linear combination by central Wishart distributions. Furthermore, in the context of multivariate Behrens-Fisher problem, a similar approximation to solve the linear sum of Wishart matrices has been given in \cite{Nel1986}. Therein the authors have approximated the sum by a single Wishart distribution by determining the associated degree of freedom and the parameter matrix. A very recent work in this direction is by one of the present authors, where exact matrix distribution has been computed for the sum of two Wishart matrices with arbitrary covariance matrices~\cite{Kumar2014}. Moreover, explicit result for the eigenvalue statistics has been worked out for the case when one of the Wishart matrices possesses covariance matrix proportional to the identity matrix. In the present work we are concerned with the eigenvalue statistics for the sum of arbitrary number of central Wishart matrices with covariance matrices proportional to the identity matrix. 

For the MIMO MAC channel, the ergodic sum rate capacity has never been obtained due to the lack of analytical results for the joint eigenvalue probability density function of sum of Wishart matrices. However, the capacity with perfect channel state information at receiver and transmitter (CSITR) sides is very well studied. With perfect CSIT and CSIR the system can be viewed as a set of parallel non interfering MIMO MACs. Thus, the ergodic capacity region can be obtained as an average of these parallel MIMO MAC capacity regions (see \cite{Goldsmith2003} and the references therein). Another approach is to obtain asymptotic results on the sum ergodic capacity of MIMO MAC channels. This can be done by considering that the number of receive antennas and the number of transmitters tend to infinity \cite{Goldsmith2003}.

In order to solve these and related challenging problems, we have proposed two distinct approaches that are presented in Section \ref{sec:Proposed Solution}. The first approach is the derivation of an exact closed-form expression for the marginal eigenvalue distribution of the sum of Wishart matrices. The main idea behind this solution is to demonstrate that the matrix resulting from the weighted sum of $K$ Wishart matrices can be rewritten as the product of a single matrix and its conjugate transpose. This resulting Wishart matrix happens to correspond to a covariance matrix which incorporates the information about the weights. Therefore, its eigenvalue distribution follows from the pre-existing knowledge about Wishart semicorrelated matrices. The derivation of this result is given in Appendices \ref{AppA} and \ref{AppB}. Our second proposed solution is to approximate the sum of $K$ independent Wishart matrices by just one equivalent Wishart matrix. This approach is based on the idea of equating the cumulants, as done in~\cite{Tan1983} for the case of general covariance matrices. We have found a simple and compact closed-form expression to determine the degrees of freedom of this equivalent Wishart matrix.

In order to show that our proposed solutions are valid, we have chosen the two MIMO multiuser scenario described before, viz. MIMO MAC and MIMO Relay. First, by considering an arbitrary set of parameters, we show that the Monte Carlo simulated eigenvalue distribution of the sum of Wishart matrices is perfectly described by our exact expression. Then, we apply  our approximation to find an equivalent Wishart matrix, and compare its eigenvalue distribution with the simulated results. The results are promising.

Moving a step forward, we present in Section \ref{section:Application} a new closed-form expression for the ergodic sum capacity. This expression takes as  input the exact eigenvalue distribution of the sum of Wishart matrices or the approximate eigenvalue distribution of the equivalent Wishart. We show that the analytical ergodic sum rate capacity matches the simulation results perfectly. All these results are shown in Section \ref{sec:Numerical Results} and give basis for our conclusions presented in Section \ref{sec:Conclusions}.

Besides all the sections mentioned above, we present  some fundamental definitions about Wishart matrices in Section \ref{sec:Preliminaries}.

\section{Preliminaries}
\label{sec:Preliminaries}

In this section we begin with the definition of complex Wishart distribution, which depends crucially on variance and degrees-of-freedom parameters. These are then used to construct the matrix model of our interest, namely the weighted sum of central Wishart matrices. This, in turn, is used in later sections for derivation of the probability density function and relevant metric for our problem.

Given a random $m_i$-dimensional non-negative definite matrix with $p_i$ degrees of freedom $\mathbf{W}_i \in \mathbb{C}^{m_i \times m_i}$. The distribution law of $\mathbf{W}_i$,
\begin{equation}
P_{\mathbf{W}_i}(\mathbf{W}_i)\propto \det(\mathbf{W}_i)^{p_i-m_i}\exp\left(-\tr \mathbf{\Sigma}_i^{-1} \mathbf{W}_i\right),
\end{equation}
is called \textit{complex central Wishart distribution} \cite{James1960,Goodman1963}, and is denoted by
\begin{equation}
\mathbf{W}_i \sim \mathcal{CW}_{m_i}(p_i,\mathbf{\Sigma}_i).
\label{Wishartdistriution}
\end{equation}
Here, $\mathbf{\Sigma}_i$ is the covariance matrix, $\det(\cdot)$ and tr$(\cdot)$ represent determinant and trace operators, respectively. In the following we will consider $\mathbf{\Sigma}_i=\sigma_i^2 \mathbf{I}_{m_i}$, where $\mathbf{I}_{m_i}$ is the identity matrix of dimension $m_i$.

Consider $K$ independent matrices with the distribution given by \eqref{Wishartdistriution}. We are interested in the eigenvalue statistics of the weighted sum of these $K$ matrices normalized by their respective degrees of freedom, viz.,
\begin{equation}
\overline{\mathbf{W}}=\sum_{i=1}^K \frac{a_i}{p_i}\mathbf{W}_i
\label{eq:sum:wishart}
\end{equation}
where $a_i \in \mathbb{R}^+$. Note that the above sum is possible only if the $m_i$'s are identical, say $m$.

It is known for the general case of $\overline{\mathbf{W}}=\sum_{i=1}^K \mathbf{W}_i$, with $\mathbf{W}_i \sim \mathcal{W}_{m_i}(p_i,\mathbf{\Sigma})$, that $\overline{\mathbf{W}} \sim \mathcal{W}_{m_i}(\sum_{i=1}^K p_i,\mathbf{\Sigma})$; see\cite[Theorem 7.3.2.]{anderson1958}. On the other hand, if the covariance matrices $\mathbf{\Sigma}_i$'s are not proportional to identity matrix, then obtaining the distribution of $\overline{\mathbf{W}}$ and its eigenvalues is nontrivial; see for example \cite{Kumar2014}. However, if  $\mathbf{\Sigma}_i$'s are proportional to identity matrix, then as shown in appendix \ref{AppA},  $\overline{\mathbf{W}}$ actually corresponds to a semicorrelated Wishart distributed case \cite{Ivrlac2003}.

For the scenario $\mathbf{\Sigma}_i \propto\mathbf{I}_m$, without loss of any generality we may consider $\mathbf{\Sigma}_i=\sigma^2\mathbf{I}_m$, as different $\sigma_i$ values can be absorbed in $a_i$\footnote{$\sigma$ may also be absorbed in $a_i$. Therefore, $\mathbf{W}_i$ corresponds essentially to an \emph{uncorrelated} Wishart case.}. Let us define 
\begin{align}
v_i&=(a_i/p_i)\sigma^2,\ \ i=1,...,K;\\
p&=\sum_{i=1}^K p_i.
\label{eq:vi}
\end{align}

With these definitions we now present the exact as well as approximate solution concerning the eigenvalue statistics of $\overline{\mathbf{W}}$.

\section{Proposed Solution}
\label{sec:Proposed Solution}

This section presents our contribution to determine the eigenvalue distribution for the weighted sum of $K$ Wishart matrices, as defined above. First, we present a new exact closed-form expression. Then, we propose an approximation that replaces the weighted sum of $K$ Wishart matrices by an equivalent matrix.

\subsection{Exact closed-form expression for the marginal eigenvalue distribution}

The \textbf{main} result of our paper is given in the following proposition.

\emph{Proposition 1:} The marginal density of eigenvalues of $\overline{\mathbf{W}}$ defined in \eqref{eq:sum:wishart} is given by
\begin{align}
\label{PL}
P_\lambda(\lambda)=c\det\begin{bmatrix}
0 & \left[f_{j_1}(v_1,\lambda)\right]_{j_1=1,...,p_1} & \cdots & \left[f_{j_K}(v_K,\lambda)\right]_{j_K=1,...,p_K}\\
\left[g_i(\lambda)\right]_{i=1,..,p}&  \left[h_{i_1,j_1}(v_1)\right]_{\substack{i_1=1,...,p\\j_1=1,...,p_1}} & \cdots & \left[h_{i_K,j_K}(v_K)\right]_{\substack{i_K=1,...,p\\j_K=1,...,p_K}}
\end{bmatrix}.
\end{align}

The entries $f_j(v,\lambda), g_i(\lambda), h_{i,j}(v)$ inside the determinant in the above expression are respectively given by 
\begin{align}
\label{f1}
&f_j(v,\lambda)=\Gamma(j)\,v^{j-m-1}\,\exp(-\lambda/v)\,L_{j-1}^{(m-j+1)}(\lambda/v),\\ 
\label{g1}
&g_i(\lambda)=\lambda^{m-i}/\Gamma(m-i+1),\\
\label{h1}
&h_{i,j}(v)=\frac{\Gamma(i)}{\Gamma(i-j+1)}\,v^{j-i}.
\end{align}
Here $\Gamma(\cdot)$ is the Gamma function given by $\Gamma(z)=\int_{0}^{\infty} t^{z-1} e^{-t}\ dt$, and $L_\mu^{(\nu)}(x)$ are the associated Laguerre polynomials \cite[eq. 22.5.38]{abramowitz2012handbook}.
The normalization $c$ in~\eqref{PL} is obtained using
\begin{align}
c^{-1}&=-m \det\Big[ \left[h_{i_1,j_1}(v_1)\right]_{\substack{i_1=1,...,p\\j_1=1,...,p_1}}~~~\cdots~~~ \left[h_{i_K,j_K}(v_K)\right]_{\substack{i_K=1,...,p\\j_K=1,...,p_K}}\Big].
\end{align}

\emph{Proof:} See Appendix A.

\subsection{Approximation for Sum of Wishart matrices}

\emph{Proposition 2:} Given the weighted sum of $K$ Wishart matrices normalized by their respective degrees of freedom as \eqref{eq:sum:wishart}, we propose the following approximation
\begin{equation}
\overline{\mathbf{W}}\approx \mathbf{S}\dfrac{\left( \sum^K_{i=1} a_i \right)}{p_s} ,
\label{eq:Wexpec2}
\end{equation}
where $\mathbf{S} \sim \mathcal{CW}_{m_s}(p_s,\mathbf{\Sigma}_s)$, $\mathbf{\Sigma}_s=\sigma^2 \mathbf{I}_{m}$, and $p_s$ given by
\begin{align}
p_s = \left\lfloor \frac{ \left( \sum^K_{i=1} a_i\right)^2  }{\sum^K_{i=1} \frac{a_i^2}{p_i}} \right\rceil,
\label{eq:ps}
\end{align}
with $\lfloor \cdot \rceil$ representing the nearest integer operator.

\textit{Proof:} The rationality for the approximation is as follow. The expected value of $\mathbf{W}_i$ given in \eqref{Wishartdistriution} is given by \cite{Goodman1963,Maiwald1997}
\begin{equation}
\mathbb{E}[\mathbf{W}_i] = p_i\mathbf{\Sigma}_i,
\end{equation}
The variance of the main diagonal elements are given by \cite{Maiwald1997}
\begin{align}
\mbox{var}[\mathbf{W}_i(j,j)] = p_i\, \sigma^4,
\end{align}
where we have dropped the subscript of $\sigma$ as explained before.

Also, the expected value of $\overline{\mathbf{W}}$ in \eqref{eq:sum:wishart} is given by
\begin{align}
\mathbb{E}\left[\overline{\mathbf{W}}\right] = \mathbb{E}\left[\sum^K_{i=1} \frac{a_i}{p_i} \mathbf{W}_i \right] = \sum^K_{i=1} a_i\frac{p_i}{p_i} \mathbf{\Sigma}_i =  \sum^K_{i=1} a_i \mathbf{\Sigma}_i=\sigma^2\mathbf{I}_m\sum^K_{i=1} a_i,
\label{eq:Wexpec}
\end{align}
and the variance of the main diagonal elements is given by
\begin{align}
\mbox{var}[\overline{\mathbf{W}}(j,j)] & =  \mbox{var}\left[\sum^K_{i=1} \frac{a_i}{p_i} \mathbf{W}_i(j,j)\right]
 = \sum^K_{i=1} \frac{a_i^2}{p_i^2} p_i \sigma^4 
 =  \sigma^4 \sum^K_{i=1} \frac{a_i^2}{p_i}.
\label{eq:var1}
\end{align}

Notice that the expectation value of RHS of \eqref{eq:Wexpec2} is given by
\begin{equation}
\mathbb{E}\left[\frac{\left( \sum^K_{i=1} a_i \right)}{p_s} \mathbf{S}\right] = \left( \sum^K_{i=1} a_i \right) \frac{p_s}{p_s} \mathbf{\Sigma}_s= \sigma^2\mathbf{I}_m \sum^K_{i=1} a_i ,
\label{eq:Sexpec}
\end{equation}
and the variance of the main diagonal elements is given by
\begin{align}
\mbox{var}\left[ \frac{\left( \sum^K_{i=1} a_i \right)}{p_s} \mathbf{S}(j,j) \right] & = \sigma^4 p_s \frac{\left( \sum^K_{i=1} a_i\right)^2}{p_s^2} 
=  \sigma^4 \frac{\left( \sum^K_{i=1} a_i\right)^2}{p_s}.
\label{eq:var2}
\end{align}

Here, we call attention to the similarities of results of \eqref{eq:Wexpec} and \eqref{eq:Sexpec}, as well as \eqref{eq:var1} and \eqref{eq:var2}. Therefore, it is possible to state a relation between the degrees of freedom of different Wishart distributions by equating \eqref{eq:var1} and \eqref{eq:var2} and obtaining the closed-form expression for $p_s$ given by \eqref{eq:ps}. Also, since $p_s$ is related with the number of columns of the $\mathbf{S}$, it should be an integer number, and this is why we have to apply the nearest integer operation in \eqref{eq:ps}.

\section{Application}
\label{section:Application}

Generally, a single user communication under fading conditions has a received signal expression given by \cite[(5.86)]{tse2005fundamentals}
\begin{align}
\mathbf{y}=h \mathbf{x} + \mathbf{z}
\end{align}
where $\mathbf{z} \sim \mathcal{CN}\left(0,1\right)$ is the noise, $\mathbf{x}$ is the Gaussian distributed input signal with power constraint $\lVert\mathbf{x}\rVert^2\leq a_i$, and $h$ is the channel gain. Herein, we assume $h \sim \mathcal{CN}\left(0,\sigma^2\right)$, therefore, the channel is under Rayleigh fading. Now, suppose that the source has $M_i$ transmitting antennas, and the destination has $N_i$ receiving antennas. Hence, the wireless channel is described by the complex $N_i$ by $M_i$ random matrix $\mathbf{H}_i$, and the received signal is given by \cite[(7.1)]{tse2005fundamentals}
\begin{align}
\mathbf{y}=\mathbf{H}_i \mathbf{x} + \mathbf{z}
\label{y:MIMO:PTP}
\end{align}
where $\mathbf{z} \sim \mathcal{CN}\left(0,\mathbf{I}_{m_i}\right)$ is the white Gaussian noise vector at a symbol time (not show here by simplicity), $\mathbf{x} \in \mathcal{CN}^{M_i}$, and $\mathbf{y} \in \mathcal{CN}^{N_i}$. The entries of $\mathbf{H}_i$ are $h_{rt}$, with $1 \leq r \leq N_i$ and $1 \leq t \leq M_i$. Define a matrix $\mathbf{W}_i$ as \cite{telatar1999capacity}
\begin{equation}
\mathbf{W}_i=\left\{\begin{array}{cc}
\mathbf{H}_i\mathbf{H}_i^\dagger\ \ N_i < M_i\\
\mathbf{H}_i^\dagger\mathbf{H}_i\ \ N_i \geq M_i,
\end{array}\right.
\label{Wishart}
\end{equation}
where $\dagger$ denotes the transpose conjugated matrix operator. Hence, $\mathbf{W}_i$ has real, non-negative eigenvalues  \cite{James1964}. The matrix $\mathbf{
W}_i$ is distributed as \eqref{Wishartdistriution} with $p_i=\mbox{max}(M_i,N_i)$ and $m_i=\mbox{min}(M_i,N_i)$.

Telatar has shown in his canonical paper \cite{telatar1999capacity}, that the ergodic capacity for the system described by \eqref{y:MIMO:PTP} is given by
\begin{align}
\mathcal{C}=\mathbb{E}_{\mathbf{W}_i}\left[\log_2 \det\left(\mathbf{I}_{m_i}+\frac{a_i}{p_i}\mathbf{W}_i \right)\right]=m_i\int_0^\infty  \log_2(1+\lambda)P_\lambda(\lambda)\ d\lambda.
\label{Cerg:MIMO:PTP}
\end{align}
where $P_\lambda(\lambda)$ is the the marginal density of eigenvalues. 
Since in this work we are interested in multiuser scenario instead of a single user described above, we should adapt the capacity equations for a multiuser case.

\begin{figure}
\centering
\includegraphics[scale=0.31]{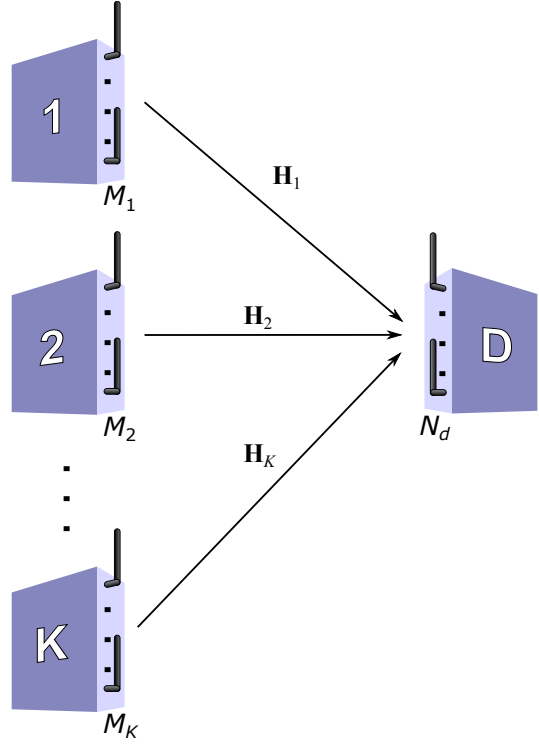}
\caption{System Model of MIMO MAC channel. The  destination (D), equipped with $N_d$ receiving antennas, receives signal from $K$ sources, each one equipped with $M_i$ transmitting antennas, with  $i=1,\dots,K$.}
\label{MAC_Channel_model}
\end{figure}

The first scenario is the MIMO MAC depicted in Fig.~\ref{MAC_Channel_model}. The channel capacity for a MIMO MAC network with $K$ sources and one destination under Rayleigh fading is given by \cite{Goldsmith2003}
\begin{align}
\mathcal{C}_\text{MAC}&=\mathbb{E}_{\mathbf{W}_i}\left[\log_2 \det\left(\mathbf{I}_{m}+ \sum_{i=1}^K \frac{a_i}{p_i}\mathbf{W}_{i}\right)\right]\nonumber\\
&=\mathbb{E}_{\overline{\mathbf{W}}}\left[\log_2 \det\left(\mathbf{I}_{m}+ \overline{\mathbf{W}}\right)\right]\nonumber\\
&=\mathbb{E}_{\overline{\mathbf{W}}}\left[\tr \log_2 \left(\mathbf{I}_{m}+ \overline{\mathbf{W}}\right)\right]
\label{eq:Capacity}
\end{align}
where we have used first the equality given in \eqref{eq:sum:wishart}, and then the property of matrices that asserts that $\det(\exp(\mathbf{A}))=\exp(\tr(\mathbf{A}))$ \cite{hornMatrix}.

In order to solve $\mathcal{C}_\text{MAC}$ given in \eqref{eq:Capacity}, we can use the eigenvalue distribution given in \eqref{PL} to obtain a closed-form expression for the ergodic sum rate capacity given in \eqref{Cav}  with the appropriated parameters of the Wishart matrices given in the problem statement. In terms of the marginal density $P_\lambda(\lambda)$ of eigenvalues of $\overline{\mathbf{W}}$, we can write
\begin{equation}
\label{C}
\mathcal{C}_\text{MAC}=m\int_0^\infty P_\lambda(\lambda)\log_2(1+\lambda)\ d\lambda.
\end{equation}
As shown in the appendices, an exact closed form expression for the ergodic capacity can be obtained in determinantal form involving Meijer-G functions, and is given by
\begin{equation}
\label{Cav}
\mathcal{C}_\text{MAC}=-m\,c\, \sum_{\mu=1}^m\det 
\begin{bmatrix}
[\psi_{i_1,j_1}^{(\mu)}(v_1)]_{\substack{i_1=1,...,p\\j_1=1,...,p_1}} & ... & [\psi_{i_K,j_K}^{(\mu)}(v_K)]_{\substack{i_K=1,...,p\\j_K=1,...,p_K}}
\end{bmatrix},
\end{equation}
where $\psi_{i,j}^{(\mu)}(v)$ is given by
\begin{align}
\label{psi1}
 \psi_{i,j}^{(\mu)}(v)=
\begin{cases}
\mathcal{G}_{i,j}(v), & i=\mu, \\ 
h_{i,j}(v), & i\neq \mu,
 \end{cases}
\end{align}
with
\begin{equation}
\label{G1}
\mathcal{G}_{i,j}(v)=\dfrac{v^{j-1}}{(\ln 2)\,\Gamma(m-i+1)}\MGG{0}{i-1}{i}{i-1}{i-1}{m}{j-1}{\dfrac{1}{v }},
\end{equation}
and $h_{i,j}(v)$ as in~\eqref{h1}.

On the other hand, by using the proposed approximation given in \eqref{eq:Wexpec2}, \eqref{eq:Capacity} becomes
\begin{align}\label{eq:capacity:eig}
\mathcal{C}_\text{MAC}=\mathbb{E}_{\mathbf{W}_i}\left[\log_2 \det\left(\mathbf{I}_{m}+ \sum_{i=1}^K \frac{a_i}{p_i}\mathbf{W}_{i}\right)\right]\nonumber\\
\approx\mathbb{E}_{\mathbf{S}}\left[\log_2 \det\left(\mathbf{I}_{m}+ \frac{\sum^K_{i=1} a_i}{p_s} \mathbf{S}\right)\right]
\end{align}
Hence, we can use \eqref{Cav} with $p_1=p_s$ and $a_1=\sum_{i=1}^K a_i$, and then set $K=1$ to obtain $\mathcal{C}_\text{MAC}$.

\begin{figure}
\centering
\includegraphics[scale=0.31]{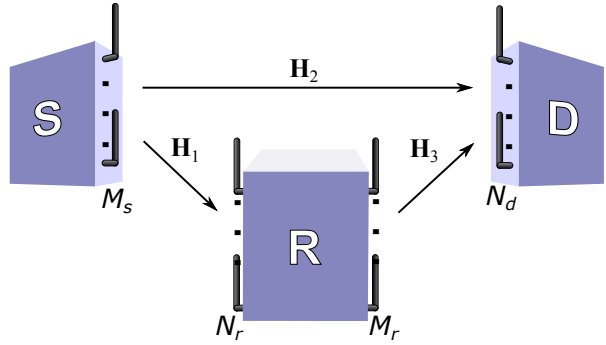}
\caption{System Model of MIMO relay channel. The source (S), equipped with $M_s$ transmitting antennas, wants to communicate with the destination (D), equipped with $N_d$ receiving antennas. The relay (R), equipped with $M_r$ and $N_r$ transmitting and receiving antennas, respectively, could collaborate in this communication.}
\label{MIMO_Channel_model}
\end{figure}

Now, let's turn our attention to the MIMO Relay channel shown in Fig.~\ref{MIMO_Channel_model}. This channel can be viewed as a composition of a MAC and broadcast channel (BC) with $K=2$. Suppose that the channel gains are known at the corresponding receivers only (CSI). In this scenario, an upper bound on the ergodic capacity of the MIMO relay channel is given by \cite[Theorem 4.1]{Host-Madsen2005}
\begin{equation}
\mathcal{C}_\text{upper}=\min (\mathcal{C}_\text{BC},\mathcal{C}_\text{MAC}) 
\end{equation}
where
\begin{align}
\mathcal{C}_\text{BC}=\mathbb{E}_{\mathbf{W}_i}\left[\log_2 \det\left(\mathbf{I}_{m_i}+ \frac{a_1}{p_1}\mathbf{W}_{1} +\frac{a_2}{p_2}\mathbf{W}_{2} \right)\right],
\label{eq:Capacity:BC}
\end{align}
and $\mathcal{C}_\text{MAC}$ as in \eqref{eq:Capacity}. Notice that $\mathcal{C}_\text{BC}$ is similar to $\mathcal{C}_\text{MAC}$ given in \eqref{eq:Capacity}. Therefore a similar procedure can be implemented to obtain $\mathcal{C}_\text{BC}$.

\section{Numerical Results}
\label{sec:Numerical Results}

In this section we have obtained numerical results for the closed form expressions and for the proposed approximation. The results are compared with Monte Carlo simulations to validate the analytical expressions. For each one of the simulations, 40,000 channel realizations were performed. In all cases, there is a perfect agreement between analytical and simulation results. We have chosen three arbitrary scenarios and one well-known scenario from \cite{Host-Madsen2005}.

\begin{table}
\caption{Simulation Parameters and Results for $\mathcal{C}$}
\centering
\begin{tabular}{|c|c|c|c|c|c|c|}
\hline
\multirow{2}{*}{Case} & \multirow{2}{*}{$a_i$ (dB)} & \multirow{2}{*}{MIMO} & \multirow{2}{*}{$p_s$} & \multicolumn{3}{|c|}{Ergodic Sum Rate Capacity (bits)} \\ \cline{5-7}
& & & &Simulation & Analytical & Approximation \\
\hline
\multirow{5}{*}{I} & $a_1=19.8$ & \multirow{5}{*}{$4 \times 4$} & \multirow{5}{*}{13} & \multirow{5}{*}{44.20} & \multirow{5}{*}{44.20} & \multirow{5}{*}{44.15} \\
& $a_2=29.5$& & & & &\\ 
& $a_3=29.8$& & & & &\\
& $a_4=26.1$& & & & &\\
& $a_5=21.7$& & & & &\\ 
\hline
\multirow{10}{*}{II} & $a_1=28.3$ & \multirow{10}{*}{$8 \times 8$} & \multirow{10}{*}{51} & \multirow{10}{*}{93.86} & \multirow{10}{*}{93.86} & \multirow{10}{*}{93.85} \\
& $a_2=17.7$& & & & &\\ 
& $a_3=26.5$& & & & &\\
& $a_4=27.3$& & & & &\\
& $a_5=29.3$& & & & &\\ 
& $a_6=21.5$& & & & &\\ 
& $a_7=19.5$& & & & &\\ 
& $a_8=27.9$& & & & &\\ 
& $a_9=9.3$& & & & &\\ 
& $a_{10}=24.3$& & & & &\\ 
\hline
\multirow{3}{*}{III} & $a_1=9.7$ & \multirow{3}{*}{$2\times2$}& \multirow{3}{*}{$3-4$} & \multirow{3}{*}{10.94} & \multirow{3}{*}{10.94} & \multirow{3}{*}{11.01} \\
& $a_2=17.6$& & & & &\\ 
& $a_3=16.7$& & & & &\\
\hline
\end{tabular}
\label{tab:simres}
\end{table}

\begin{figure}
\centering
\includegraphics[scale=.9]{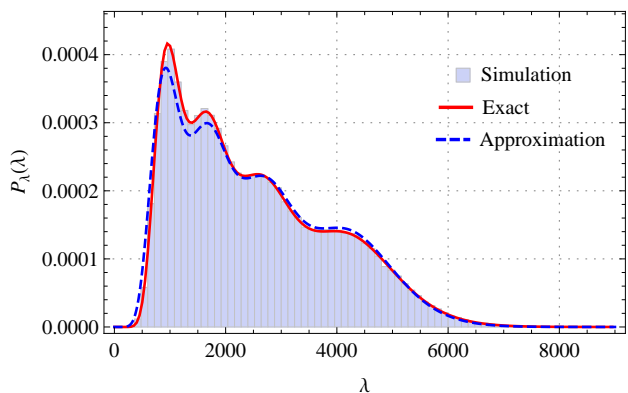}
\caption{Marginal distribution of eigenvalues of $\overline{\mathbf{W}}$ for Case I (see Table \ref{tab:simres}). Simulation results are in perfect agreement with the closed form analytical distribution (red line). The distribution for the proposed approximation, shown with blue dashed line, is also very close to the exact result.}
\label{fig:SIM01}
\end{figure}

Consider a MIMO MAC scenario shown in Fig.~\ref{MAC_Channel_model} with $K=5$ users, each one with $M_i=4$ transmitting antennas, where $i=1,\dots,K$. Destination node D has $N_d=4$ receiving antennas. The normalized signal to noise ratios $(a_i)$ at destination were arbitrarily chosen, and are shown in Table \ref{tab:simres} (Case I). The marginal eigenvalue distribution is shown in Fig.~\ref{fig:SIM01}. Notice the perfect agreement between the simulation and analytical results. The approximate ergodic sum rate capacity was also computed using the equivalent matrix $\mathbf{S}$ (with $p_s=13$ degrees of freedom) and is shown in the last column of Table \ref{tab:simres}. The eigenvalue distribution for the approximation is also shown in Fig.~\ref{fig:SIM01}. Note  how close the approximation and the ergodic sum rate capacity are to the exact values.

\begin{figure}
\centering
\includegraphics[scale=.95]{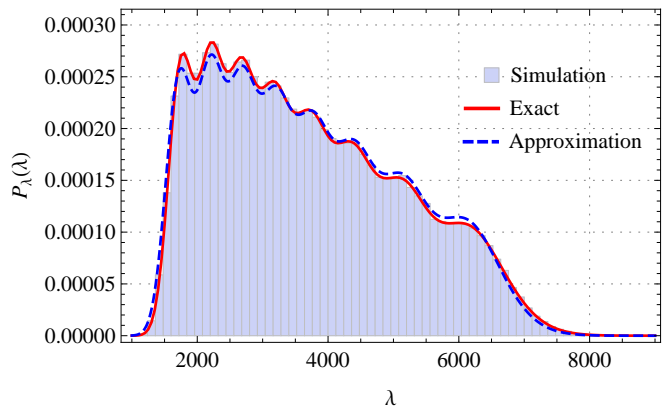}
\caption{Marginal distribution of eigenvalues of $\overline{\mathbf{W}}$ for Case II (see Table \ref{tab:simres}). Simulation results match perfectly with the closed form analytical distribution (red line). The distribution for the proposed approximation is also shown using blue line.}
\label{fig:SIM02}
\end{figure}

In the next scenario we have increased the number of users to $K=10$ with MIMO $8 \times 8$. The $a_i$ coefficients are given in Case II of Table \ref{tab:simres}. Notice again in Fig.~\ref{fig:SIM02} a perfect match between analytical and simulation results. The ergodic capacity results also agree, as can be seen in Table \ref{tab:simres}. The eigenvalue distribution from the approximation is shown in Fig.~\ref{fig:SIM02} as well.

\begin{figure}
\centering
\includegraphics[scale=.95]{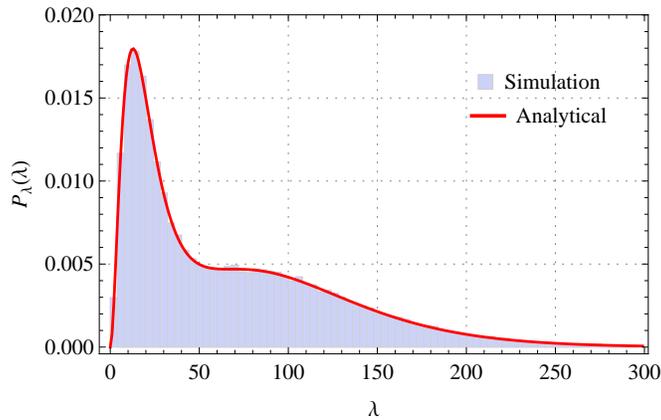}
\caption{Marginal distribution of eigenvalues of $\overline{\mathbf{W}}$ for Case III (BC). Simulation results are in perfect match with the closed form analytical distribution (line).}
\label{fig:SIM03}
\end{figure}

\begin{figure}
\centering
\includegraphics[scale=.95]{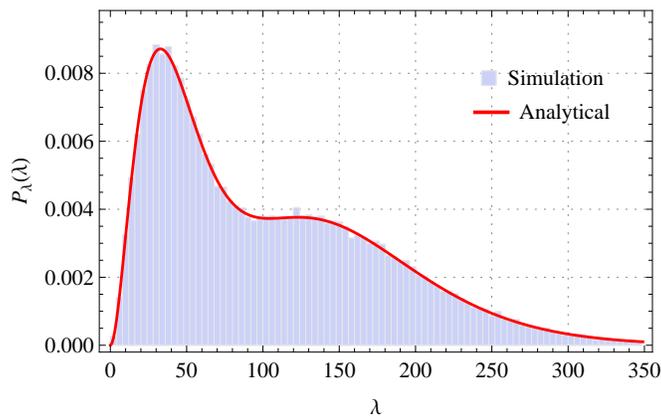}
\caption{Marginal distribution of eigenvalues of $\overline{\mathbf{W}}$ for Case III (MAC). Simulation results match perfectly with the closed form analytical distribution (line).}
\label{fig:SIM04}
\end{figure}

Now let us move on to a more involved scenario depicted in Fig.~\ref{MIMO_Channel_model}. The ergodic capacity was originally calculated in \cite{Host-Madsen2005} using convex programming. The parameters used are depicted in Case III of Table \ref{tab:simres} and the plot for the eigenvalue distribution  is given in Fig.~\ref{fig:SIM03}. Fig.~\ref{fig:SIM04} shows the eigenvalue distribution for the MAC channel with the following parameters $\left(a_2=17.6,\ a_3=16.7\right)$. The upper bound on ergodic capacity, as mentioned before, is the minimum of the capacity of BC and MAC, as given in Table \ref{tab:simres}.

\begin{figure}
\centering
\includegraphics[scale=.9]{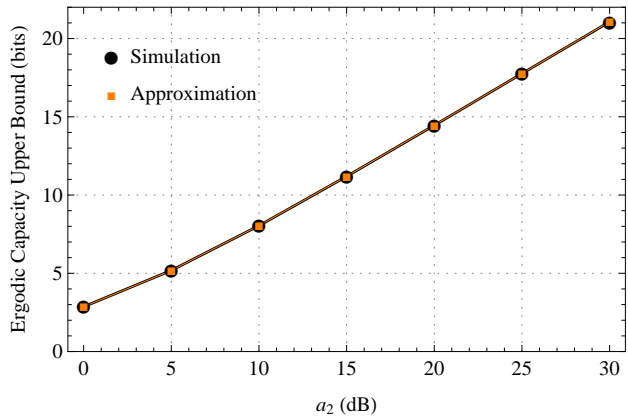}
\caption{Ergodic Capacity Upper Bound versus $a_2$ with constraints $a_3=a_2$ and $a_1=10 a_2$. Based on \cite[Fig. 5]{Host-Madsen2005}. The simulation results are in perfect agreement with the analytical results obtained with the proposed approximation.}
\label{Cap_vs_SNR_MIMO_Relay}
\end{figure}

Besides the Case III, we have also reproduced the scenario given in \cite[Fig. 5]{Host-Madsen2005} for MIMO Relay channel. This scenario is well known because the upper bound and the lower bound ``converge''. \emph{That is to say, the ergodic capacity of the MIMO relay channel over Rayleigh fading can be characterized under this SNR condition} \cite{Host-Madsen2005}. The constraints for this scenario are $a_2=a_3$, $a_1=10a_2$, and $0 \leq a_2 \leq 30$ dB. The ergodic capacity results for Monte Carlo simulation and the proposed approximation are given in Fig.~\ref{Cap_vs_SNR_MIMO_Relay}. Notice again the perfect agreement of the results.

\begin{figure}
\centering
\includegraphics[scale=.9]{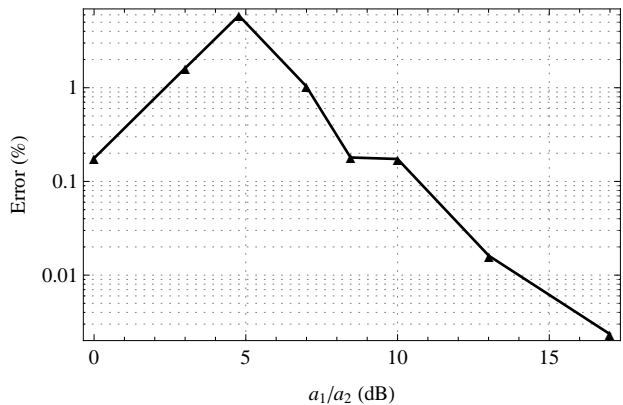}
\caption{Error of approximation versus weight ratio $a_1/a_2$ and $a_2=5$.}
\label{Approximation_Error}
\end{figure}

The results in Table \ref{tab:simres} and in Fig.~\ref{Cap_vs_SNR_MIMO_Relay} show that the proposed approximation results in a very small difference from the exact result for all mentioned scenarios. Since the approximation depends on the weight factor of each matrix, or in other words depends on the SNR of each channel, it would be interesting to investigate cases where these weight factors varies. Fig.~\ref{Approximation_Error} shows the percentage error as the ratio $a_1/a_2$ varies from 0 to 15 dB. Note that the error is less than 1$\%$ for $0<a_1/a_2<2$ and $a_1/a_2>7$.

\section{Conclusions}
\label{sec:Conclusions}

In this work we uncovered a one to one correspondence between the weighted sum of arbitrary number of \emph {uncorrelated} central Wishart matrices and a single semicorrelated Wishart matrix. Using this observation we presented a closed form expression for the marginal distribution of the eigenvalues for the weighted sum of $K$ complex central Wishart matrices. To the best of our knowledge this problem has not been tackled before. Here the motivation for establishing a result emerged from the multiuser information theory area. However, since Wishart matrices play crucial role in diverse fields, we believe that our results are relevant to these as well. We would like to remark that it is also possible to obtain results for the joint probability density of all eigenvalues, and correlation functions involving distribution of two or more eigenvalues.

We applied our new closed-form expression for analyzing the ergodic sum rate capacity of MIMO multiuser channels. Moreover, we also derived a closed form expression for the ergodic channel capacity and used it to obtain the capacities for MIMO MAC and MIMO Relay channel. Besides the closed form exact expression for marginal distribution, we also proposed an approximation that is very simple and presents promising results when used to obtain ergodic sum capacity. In addition, we confirmed the validity of all our analytical expressions by Monte Carlo simulations.

\appendices

\section{Mapping to a semicorrelated Wishart distribution}
\label{AppA}

Consider $p_i\times m$ dimensional complex matrices $\mathbf{H}_i$, $i=1,...,K$, from the normal distribution:
\begin{equation}
P_i(\mathbf{H}_i)\propto\exp\left(\tr \mathbf{H}_i {\mathbf{\Sigma}_i}^{-1} \mathbf{H}_i^\dag\right),
\end{equation}
where $\mathbf{\Sigma}_i=\sigma^2\mathbf{I}_{m}$.
Then $m\times m$-dimensional matrices ${\mathbf{W}}_i=\mathbf{H}_i^\dag \mathbf{H}_i$, are respectively from the complex-Wishart distribution given in \eqref{Wishartdistriution}. The matrix $\overline{\mathbf{W}}$ can be written as
\begin{align}
\nonumber
\overline{\mathbf{W}}&=\sum_{i=1}^K \frac{a_i}{p_i}\mathbf{H}_i^\dag \mathbf{H}_i=\sum_{i=1}^K \mathbf{H}_i^\dag\left(\frac{a_i}{p_i}\mathbf{I}_{p_i}\right) \mathbf{H}_i\\
&=\begin{bmatrix}\mathbf{H}_1^\dag & \mathbf{H}_2^\dag & \hdots & \mathbf{H}_K^\dag \end{bmatrix}
\begin{bmatrix}\dfrac{a_1}{p_1}\mathbf{I}_{p_1} & 0 &\cdots & 0\\
\nonumber
0 & \dfrac{a_2}{p_2}\mathbf{I}_{p_2} & \cdots & 0\\
\vdots & &\ddots &\vdots\\
0 & 0 &\cdots & \dfrac{a_K}{p_K}\mathbf{I}_{p_K}\\
 \end{bmatrix}
\begin{bmatrix}\mathbf{H}_1 \\ \mathbf{H}_2 \\ \vdots \\ \mathbf{H}_K \end{bmatrix}\\
&=\mathbf{H}^\dag {\bf D} \mathbf{H} \equiv \mathbf{G}^\dag \mathbf{G}.
\end{align}
We defined here
\begin{align}
&\mathbf{H}^\dag=\begin{bmatrix}\mathbf{H}_1^\dag & \mathbf{H}_2^\dag & \hdots & \mathbf{H}_K^\dag \end{bmatrix},\\
&{\bf D}=\diag\left[\dfrac{a_1}{p_1}\mathbf{I}_{p_1},\cdots,\dfrac{a_K}{p_K}\mathbf{I}_{p_K}\right],\\
& \mathbf{G}={\bf D}^{1/2} \mathbf{H}.
\end{align}
With the above information it is clear that $\mathbf{G}$ satisfies the distribution
\begin{equation}
\label{PG}
P_\mathbf{G}(\mathbf{G})\propto\,\exp[-\tr (\mathbf{G}^\dag{\bf V}^{-1} \mathbf{G})],
\end{equation} 
where ${\bf V}=\diag[v_1\mathbf{I}_{p_1},...,v_K\mathbf{I}_{p_K}]$, with $v_i$ (and $p$) as defined in \eqref{eq:vi}. Therefore, we are looking essentially at a semicorrelated Wishart case, with the diagonal-covariance matrix possessing some equal-value entries (\emph {multiplicities/degeneracies}). Thus, the problem boils down to determining the eigenvalue statistics of $\mathbf{G}^\dag \mathbf{G}$.
We start with the case of a diagonal covariance matrix with unequal entries along the diagonal, i.e., we will use $\hat{\bf V}=\diag(\hat{v}_1,...,\hat{v}_{p})$ in the above distribution instead of ${\bf V}$, work out the results for this case, and eventually take adequate limits to obtain the case of ${\bf V}$.

For the semicorrelated Wishart matrices, exact result for the marginal density of eigenvalues is available from several notable works \cite{Alfano2004,Simon2006,Recher2010,Recher2012}. We use here the form derived in~\cite{Recher2010,Recher2012}. Consider $p\times m$ dimensional complex matrices $\mathbf{G}$ taken from the distribution~\eqref{PG}, but with the covariance matrix $\hat{\bf V}$ as defined above. The marginal density of $p$ eigenvalues of $\mathbf{G}\mathbf{G}^\dag $, for $p\leq m$, is given to be
\begin{equation}
\label{PL1}
P_\lambda(\lambda)=-\frac{1}{p\,\Delta_p(\{\hat{v}^{-1}\}) }
\det\begin{bmatrix} 
0 & \left[\frac{\exp(-\lambda/\hat{v}_j)}{\hat{v}_j^m}\right]_{j=1,...,p} \\
\left[\frac{\lambda^{m-i}}{\Gamma(m-i+1)}\right]_{i=1,...,p} & \left[\hat{v}_j^{-i+1}\right]_{i,j=1,...,p}
\end{bmatrix}.
\end{equation}
Here $\Delta_p(\{\hat{v}^{-1}\})=\det[\hat{v}_j^{-i+1}]=\prod_{i>j}(\hat{v}_i^{-1}-\hat{v}_j^{-1})$ is the Vandermonde determinant. Since $\mathbf{G}^\dag \mathbf{G}$ and $\mathbf{G}\mathbf{G}^\dag$ share the same nonzero eigenvalues, the above result holds for $p>m$ as well, with the factor $p$ appearing in the denominator of the prefactor replaced by $m$. Moreover, in this case the bottom $p-m$ entries of the first column in the determinant comprise diverging gamma function in the denominator, and hence become zero.

We now use the relation \eqref{Cerg:MIMO:PTP} to derive the ergodic channel capacity. To this end we expand \eqref{PL1} using the first column and obtain
\begin{equation}
P_\lambda(\lambda)=-\frac{1}{m\,\Delta_p(\{\hat{v}^{-1}\}) }\, \sum_{\mu=1}^m (-1)^{\mu}\frac{\lambda^{m-\mu}}{\Gamma(m-\mu+1)}\det\begin{bmatrix} \left[\frac{\exp(-\lambda/\hat{v}_j)}{\hat{v}_j^m}\right]_{j=1,...,p} \\ \left[\hat{v}_j^{-i+1}\right]_{\substack{i,j=1,...,p \\ (i\neq \mu)}}\end{bmatrix} .
\end{equation}
Next we bring in the $\lambda^{m-\mu}/\Gamma(m-\mu+1)$ factors occurring before the determinants to the respective first rows, i.e., with $\exp(-\lambda/\hat{v}_j)/\hat{v}_j^m$. This gives
\begin{equation}
P_\lambda(\lambda)=-\frac{1}{m\,\Delta_p(\{\hat{v}^{-1}\}) }\, \sum_{\mu=1}^m (-1)^{\mu} \det\begin{bmatrix} \left[\frac{\lambda^{m-\mu}}{\Gamma(m-\mu+1)}\frac{\exp(-\lambda/\hat{v}_j)}{\hat{v}_j^m}\right]_{j=1,...,p} \\ \left[\hat{v}_j^{-i+1}\right]_{\substack{i,j=1,...,p \\ (i\neq \mu)}}\end{bmatrix} .
\end{equation}
The above equation serves as yet another expression for the marginal density.

For ergodic capacity we use \eqref{Cerg:MIMO:PTP} and obtain the following expression by interchanging the $\lambda$-integral and the summation:
\begin{equation}
\mathcal{C}=-\frac{1}{\,\Delta_p(\{\hat{v}^{-1}\}) } \sum_{\mu=1}^m(-1)^{\mu} \int_0^\infty d\lambda\, \left(\det\begin{bmatrix} \left[\frac{\lambda^{m-\mu}}{\Gamma(m-\mu+1)}\frac{\exp(-\lambda/\hat{v}_j)}{\hat{v}_j^m}\right]_{j=1,...,p} \\ \left[\hat{v}_j^{-i+1}\right]_{\substack{i,j=1,...,p \\ (i\neq \mu)}}\end{bmatrix} \right)\, \log_2  (1+\lambda).
\end{equation}
The $\lambda$-integral can be brought into the first row of the determinant, along with the factor $\log_2 (1+\lambda)$ to yield
\begin{equation}
\label{Capp}
\mathcal{C}=-\frac{1}{\,\Delta_p(\{\hat{v}^{-1}\})}  \sum_{\mu=1}^{m}(-1)^{\mu} \det\begin{bmatrix} \left[\mathcal{G}_{\mu,j}(\hat{v})\right]_{j=1,...,p} \\ \left[\hat{v}_j^{-i+1}\right]_{\substack{i,j=1,...,p \\ (i\neq \mu)}}\end{bmatrix},
\end{equation}
where
\begin{equation}
\mathcal{G}_{\mu,j}(\hat{v})=\int_0^\infty d\lambda\, \frac{\lambda^{m-\mu}}{\Gamma(m-\mu+1)}\frac{\exp(-\lambda/\hat{v}_j)}{\hat{v}_j^m}\log_2 (1+\lambda).
\end{equation}
This integral can be expressed in a closed form with the aid of Meijer-G function~\cite{Prudnikov1990}. This is facilitated by considering the following special cases of Meijer-G functions:
\begin{equation}
G^{1,0}_{0,1} \left(
\begin{array} {c}
\_ \\ \beta \end{array}\Bigg|\, z \right)= z^{\beta}e^{-z},~~~~~~
G^{1,2}_{2,2} \left(
\begin{array} {c}
1,\, 1 \\ 1,\, 0 \end{array}\Bigg|\, z \right)= \ln(1+z),
\end{equation}
We also use the convolution integral satisfied by Meijer-G function:
\begin{align}
\nonumber
&\int_0^\infty dz\,G^{m,n}_{p,q} \left(
\begin{array} {c}
a_1,\,\cdots\, a_p \\ b_1,\,\cdots\, b_q \end{array}\Bigg|\, \eta z \right)\,
G^{\mu,\nu}_{\sigma,\tau} \left(
\begin{array} {c}
c_1,\,\cdots\, c_\sigma \\ d_1,\,\cdots\, d_\tau \end{array}\Bigg|\, \omega z \right)\\
&=\frac{1}{\eta}\,G^{n+\mu,m+\nu}_{q+\sigma,p+\tau} \left(
\begin{array} {c}
-b_1,\cdots,-b_m,\,c_1,\,\cdots\, c_\sigma,\,-b_{m+1},\cdots,\,-b_q \\-a_1,\cdots,-a_n,\,d_1,\,\cdots\, d_\tau,\,-a_{n+1},\cdots,\,-a_p \end{array}\Bigg|\, \frac{\omega}{\eta} \right)\\
\nonumber
&=\frac{1}{\omega}\,G^{m+\nu,n+\mu}_{p+\tau,q+\sigma} \left(
\begin{array} {c}
a_1,\cdots,a_n,\,-d_1,\,\cdots\, -d_\tau,\,a_{n+1},\cdots,\,a_p \\b_1,\cdots,b_m,\,-c_1,\,\cdots\, -c_\sigma,\,b_{m+1},\cdots,\,b_q \end{array}\Bigg|\,\frac{\eta}{\omega}  \right).
\end{align}
The restrictions on the indices for this integration formula can be found in~\cite{Prudnikov1990}. Therefore, we obtain a closed form expression for $\mathcal{G}_{i,j}(\hat{v})$ as given  below in~\eqref{MG}. Afterwards we perform row interchanges in the determinants to bring $\mathcal{G}_{\mu,j}$ in the respective $\mu$th row. This leads to the removal of $(-1)^\mu$ factor. Consequently, we arrive at the following expression for ergodic channel capacity:

\begin{equation}
\label{Cav1}
\mathcal{C}=-\frac{1}{\det[\hat{v}_j^{1-i}]_{i,j=1,...,p}}\, \sum_{\mu=1}^m \det\left[ \psi_{i,j}^{(\mu)}(\hat{v})\right]_{i,j=1,...,p} .
\end{equation}
Here
\begin{align}
 \psi_{i,j}^{(\mu)}(\hat{v})=
\begin{cases}
\mathcal{G}_{i,j}(\hat{v}), & i=\mu \\ 
\hat{v}_j^{-i+1}, & i\neq \mu.
 \end{cases}
\end{align}
with
\begin{equation}
\label{MG}
\mathcal{G}_{i,j}(\hat{v})=\dfrac{1}{(\ln 2)\,\Gamma(m-i+1)}\MG{i-1}{i}{i-1}{i-1}{m}{\dfrac{1}{\hat{v}_k}}.
\end{equation}

\section{Proofs for equations~\eqref{PL} and~\eqref{Cav} }
\label{AppB}

To obtain equations~\eqref{PL} and~\eqref{Cav} we need to assign $\hat{v_1}=...=\hat{v}_{p_1}=v_1; \hat{v}_{p_1+1}=...=\hat{v}_{p_2}=v_2\,;\cdots; \hat{v}_{p_{(K-1)}+1}=...=\hat{v}_{p_K}=v_K$ in~\eqref{PL1} and~\eqref{Cav1}. However, direct substitution of these values makes the determinant in the numerator, as well as the determinant in the denominator to become zero. Therefore, we must carry out this substitution in a limiting manner, as described below.

Let us pay attention to the columns involving up to $p_1$ in~\eqref{PL1}. The ratio of the determinants appears as
\begin{equation}
\frac{\det\begin{bmatrix} 0 & (\hat{v}_1^{-1})^m e^{\hat{v}_1^{-1}\lambda} & (\hat{v}_2^{-1})^m e^{\hat{v}_2^{-1}\lambda} & \hdots &  (\hat{v}_{p_1}^{-1})^m e^{\hat{v}_{p_1}^{-1}\lambda} & \hdots & \\
\frac{\lambda^{m-i}}{\Gamma(m-i+1)} & (\hat{v}_1^{-1})^{i-1} &  (\hat{v}_2^{-1})^{i-1}  & \hdots & (\hat{v}_{p_1}^{-1})^{i-1} & \hdots &
  \end{bmatrix}}
  {\det\begin{bmatrix} 
(\hat{v}_1^{-1})^{i-1} &  (\hat{v}_2^{-1})^{i-1}  & \hdots & (\hat{v}_{p_1}^{-1})^{i-1} & \hdots &
  \end{bmatrix}}
\end{equation}
Consider for $j=2,...,p_1$, $\hat{v}_j^{-1}=\hat{v}_1^{-1}+\delta_j$ with small $\delta_j$, and Taylor-expand up to $\delta_j^{j-1}$:
$$(\hat{v}_j^{-1})^m e^{\hat{v}_j^{-1}\lambda}\approx \sum_{r=0}^{j-1}\frac{\delta_j^r}{r!}\frac{\partial^r}{\partial (\hat{v}_1^{-1})^r} (\hat{v}_1^{-1})^m e^{\hat{v}_1^{-1}\lambda}$$
$$(\hat{v}_j^{-1})^{i-1}\approx \sum_{r=0}^{j-1}\frac{\delta_j^r}{r!}\frac{\partial^r}{\partial (\hat{v}_1^{-1})^r} (\hat{v}_1^{-1})^{i-1}$$
Now, applying adequate column operations we obtain
\begin{equation}
\frac{\det\begin{bmatrix} 0 & (\hat{v}_1^{-1})^m e^{\hat{v}_1^{-1}\lambda} & \frac{\delta_j}{1!}\frac{\partial }{\partial (\hat{v}_1^{-1})} (\hat{v}_1^{-1})^m e^{\hat{v}_1^{-1}\lambda} & \hdots & \frac{\delta_j^{j-1}}{(j-1)!}\frac{\partial^{j-1}}{\partial (\hat{v}_1^{-1})^{j-1}} (\hat{v}_1^{-1})^m e^{\hat{v}_1^{-1}\lambda} & \hdots & \\
\frac{\lambda^{m-i}}{\Gamma(m-i+1)} &  (\hat{v}_1^{-1})^{i-1} &\frac{\delta_j}{1!}\frac{\partial }{\partial (\hat{v}_1^{-1})}  (\hat{v}_1^{-1})^{i-1} & \hdots & \frac{\delta_j^{j-1}}{(j-1)!}\frac{\partial^{j-1}}{\partial (\hat{v}_1^{-1})^{j-1}}(\hat{v}_1^{-1})^{i-1} & \hdots &
  \end{bmatrix}}
  {\det\begin{bmatrix} 
 (\hat{v}_1^{-1})^{i-1} &\frac{\delta_j}{1!}\frac{\partial }{\partial (\hat{v}_1^{-1})}  (\hat{v}_1^{-1})^{i-1} & \hdots & \frac{\delta_j^{j-1}}{(j-1)!}\frac{\partial^{j-1}}{\partial (\hat{v}_1^{-1})^{j-1}}(\hat{v}_1^{-1})^{i-1} & \hdots &
  \end{bmatrix}}
\end{equation}
The factors containing $\delta_j$ and factorial can be canceled out after being pulled out of the columns, both from numerator and denominator. Therefore, we are left with
\begin{equation}
\frac{\det\begin{bmatrix} 0 & (\hat{v}_1^{-1})^m e^{\hat{v}_1^{-1}\lambda} & \frac{\partial }{\partial (\hat{v}_1^{-1})} (\hat{v}_j^{-1})^m e^{\hat{v}_1^{-1}\lambda} & \hdots & \frac{\partial^{j-1}}{\partial (\hat{v}_1^{-1})^{j-1}} (\hat{v}_j^{-1})^m e^{\hat{v}_1^{-1}\lambda} & \hdots & \\
\frac{\lambda^{m-i}}{\Gamma(m-i+1)}&  (\hat{v}_1^{-1})^{i-1} &\frac{\partial }{\partial (\hat{v}_1^{-1})}  (\hat{v}_1^{-1})^{i-1} & \hdots & \frac{\partial^{j-1}}{\partial (\hat{v}_1^{-1})^{j-1}}(\hat{v}_1^{-1})^{i-1} & \hdots &
  \end{bmatrix}}
  {\det\begin{bmatrix} 
 (\hat{v}_1^{-1})^{i-1} &\frac{\partial }{\partial (\hat{v}_1^{-1})}  (\hat{v}_1^{-1})^{i-1} & \hdots & \frac{\partial^{j-1}}{\partial (\hat{v}_1^{-1})^{j-1}}(\hat{v}_1^{-1})^{i-1} & \hdots &
  \end{bmatrix}}.
\end{equation}
We have $\frac{\partial^{r-1}}{\partial (\hat{v}_1^{-1})^{r-1}}(\hat{v}_1^{-1})^{i-1}= (\Gamma(i)/\Gamma(i-r+1))(\hat{v}_1^{-1})^{i-r}$. The derivatives of $(\hat{v}_j^{-1})^m e^{\hat{v}_1^{-1}\lambda}$ can be evaluated with the aid of Rodrigues' formula for the associated Laguerre polynomials,
\begin{equation}
L_k^{(\beta)}(z)=\frac{z^{-\beta} e^z }{k!}\frac{\partial^k}{\partial z^k}\left(z^{k+\beta} e^{-z} \right),
\end{equation}
using adequate scaling of the variables. By implementing similar steps for rest of the columns, we arrive at~\eqref{PL}.

Similar steps can be used to arrive at~\eqref{Cav}, starting from~\eqref{Cav1}. The derivative of Meijer-G follows from the result
 \begin{equation}
 z^r \frac{\partial^r}{\partial z^r}\MG{a_1}{a_2}{b_1}{b_2}{b_3}{z}=\MGG{0}{a_1}{a_2}{b_1}{b_2}{b_3}{r}{z},
 \end{equation}
which is a special case of the following more general identity~\cite{Prudnikov1990}:
 \begin{equation}
 z^r \frac{\partial^r}{\partial z^r}G^{m,n}_{p,q}  \left(
\begin{array} {c}
a_1,\, \cdots,\, a_p \\ b_1,\,\cdots,\, b_q \end{array}\Bigg|\, z \right)= G^{m,n+1}_{p+1,q+1}\left(
\begin{array} {c}
0,\,a_1,\, \cdots,\, a_p \\ b_1,\,\cdots,\, b_q,\,r \end{array}\Bigg|\, z \right).
 \end{equation}



\end{document}